\documentclass[letterpaper,envcountsame,orivec]{llncs}
\usepackage{aaai}
\usepackage{times}
\usepackage{helvet}
\usepackage{courier}
\usepackage{booktabs}
\usepackage{\jobname}

\title{%
  Complexity of Propositional Abduction for Restricted Sets of Boolean Functions%
  \thanks{Supported by  ANR {\it Algorithms and complexity} 07-BLAN-0327-04 and  DFG grant VO 630/6-1.}%
}

\author{%
  Nadia Creignou \and Johannes Schmidt\\%
  Laboratoire d'Informatique Fondamentale, CNRS\\
  Universit\'{e} d'Aix-Marseille II\\%
  163, avenue de Luminy, 
  13288 Marseille Cedex 9, France
  \And %
  Michael Thomas\\%
  Institut f\"ur Theoretische~Informatik\\
  Gottfried Wilhelm Leibniz Universit\"{a}t\\%
  Appelstr.~4, 30167~Hannover, Germany 
}

\begin{document}

\maketitle

\begin{abstract}
Abduction is a fundamental and important form of non-monotonic reasoning. Given a knowledge base explaining how the world behaves it aims at finding an explanation for some observed manifestation. In this paper we focus on propositional abduction, where the knowledge base and the manifestation are represented by propositional formulae. The problem of deciding whether there exists an explanation has been shown to be \SigPtwo-complete in general. We consider variants obtained by restricting the allowed connectives in the formulae to certain sets of Boolean functions. We give a complete classification of the complexity for all considerable sets of Boolean functions. In this way, we identify easier cases, namely \NP-complete and polynomial cases; and we highlight sources of intractability. Further, we address the problem of counting the explanations and draw a complete picture for the counting complexity.
\end{abstract}

\section{Introduction}
  Abduction is a fundamental and important form of non-monotonic reasoning.
  Assume that given a certain consistent knowledge about the world, we want 
  to explain some observation. This task of finding an explanation or only 
  telling if there is one, is called abduction. Today it has many application areas spanning medical diagnosis~\cite{byaltajo89}, text
  analysis~\cite{hostapma93}, system diagnosis~\cite{stwo01},
  configuration problems~\cite{amfama02}, temporal knowledge bases~\cite{boli00} and has connections to default reasoning~\cite{selman-levesque:1990}.
  
  There are several approaches to formalize the problem of abduction.
  In this paper, we focus on \emph{logic based abduction} in which the knowledge base
  is given as a set $\Gamma$ of propositional formulae. We are interested in deciding whether there exists an \emph{explanation} $E$, \emph{i.e.},
  a set of literals consistent with $\Gamma$ such that $\Gamma$ and $E$ together entail the observation. 

  From a complexity theoretic viewpoint, the abduction problem is very hard in the sense that it is $\SigPtwo$-complete and thus situated at the
  second level of the polynomial hierarchy~\cite{eiter-gottlob:1995}. This intractability result raises the question for restrictions leading to
  fragments of lower complexity. Several such restrictions have been considered in previous works. One of the most famous amongst those is
  Schaefer's framework, where formulae are restricted to generalized conjunctive normal form with clauses from a fixed set of
  relations~\cite{crza06,noza05,noza08}.
  
  A similar yet different procedure is to rather require formulae to be constructed from a restricted set of Boolean functions $B$.
  Such formulae are called \emph{$B$-formulae}.
  This approach has first been taken by Lewis, who showed that the satisfiability problem is $\NP$-complete if and only if this set of
  Boolean functions has the ability to express the negation of implication connective $\not\limplies$~\cite{lew79}. 
  Since then, this approach has been applied to a wide range of problems including
  equivalence and implication problems~\cite{rei03,bemethvo08imp},
	satisfiability and model checking in modal and temporal logics~\cite{bhss05,bsssv07},
  default logic~\cite{bemethvo08}, and circumscription~\cite{thomas09}, among others.

  We follow this approach and show that Post's lattice allows to completely classify the complexity of propositional abduction 
  for several variants and all possible sets of allowed Boolean functions. We first examine the case where the representation of
  the manifestation is a literal.
  We show that depending on the set $B$ of allowed connectives the abduction problem is either $\SigPtwo$-complete, or $\NP$-complete, or
  in $\P$ and $\ParityL$-hard, or in  $\L$. More precisely, we prove that the complexity of this abduction problem is $\SigPtwo$-complete
  as soon as $B$ can express one of the functions $x \vee (y \wedge \neg z)$, $x \wedge (y \vee \neg z)$ or
  $(x \wedge y) \vee (x \wedge \neg z) \vee (y \wedge \neg z)$. It drops to $\NP$-complete when all functions in $B$ are monotonic and have
  the ability to express one of the functions $x \vee (y \wedge z)$, $x \wedge (y \vee z)$ or
  $(x \wedge y) \vee (x \wedge z) \vee (y \wedge z)$. The problem becomes solvable in polynomial time and is $\ParityL$-hard if $B$-formulae may depend
  on more than one variable while being representable as linear equations. Finally the complexity drops down to $\L$ in all remaining cases.

	We then examine several variants of the propositional abduction problem.
  The variants considered are obtained
  by restricting representation of the manifestation to be respectively a clause, a term or a $B$-formula. We present a complete classification in all
  cases. An overview of the results is given in Figure~\ref{fig:post's_lattice}.
  Our results highlight the sources of intractability and exhibit properties of Boolean functions that lead to an increase of the complexity
  of abduction.
  
  In \cite{crza06} the authors obtained a complexity classification of the abduction problem for formulae which are in generalized conjunctive
  normal form, with clauses from a fixed set of relations. The two classifications are in the same vein since they classify the complexity of
  abduction for local restrictions on the knowledge base. However the two results are incomparable, in the sense that no classification can be
  deduced from the other. They only overlap on the particular case of the linear connective $\xor$, for which both types of sets of formulae can be
  seen as systems of linear equations. This special abduction case has been shown to be decidable in polynomial time in \cite{zanuttini03}.
 
  Besides the decision problem, another natural question is concerned with the number of explanations. This problem refers to the counting
  problem for abduction. The study of the counting complexity of abduction has been started by Hermann and Pichler (\citeyear{hepi07}).
  We prove here a trichotomy theorem showing that counting the full explanations of propositional abduction problems is either $\SHcoNP$-complete or
  $\SHP$-complete or in $\FP$, depending on the set $B$ of allowed connectives.
 
  The rest of the paper is structured as follows. 
  We first give the necessary preliminaries. 
  Afterwards, we define the abduction problem considered herein. 
  We then classify the complexity of the abduction of a single literal. 
  These results are complemented with the complexity of the abduction problem for clauses, terms and restricted formulae.
  Next, we consider the counting problem and finally conclude with a discussion of the results.

%
	
  
\section{Preliminaries} \label{sect:prelim}
\paragraph{Complexity Theory}

  We require standard notions of complexity theory. 
  For the decision problems the arising complexity degrees encompass the classes $\L$, $\P$, $\NP$, and $\SigPtwo$. 
  For more background information, the reader is referred to~\cite{pap94}. 
  We furthermore require the class $\ParityL$ defined as the class of languages $L$ such that there exists a nondeterministic logspace Turing machine
  that exhibits an odd number of accepting paths if and only if $x\in L$, for all $x$~\cite{budaheme92}. 
  It holds that $\L \subseteq \ParityL \subseteq \P$.
  For our hardness results we consider \emph{logspace many-one reductions}, 
  defined as follows:
  a language $A$ is logspace many-one reducible to some language $B$ (written $A \leqlogm B$) if
  there exists a logspace-computable function $f$ such that $x \in A$ if and only if  $f(x) \in B$.

  A \emph{counting problem} is represented using a \emph{witness function} $w$, which for every input $x$ returns a finite set of witnesses. This witness function gives rise to the following counting problem: given an instance $x$, find the cardinality $\vert w(x)\vert$ of the witness set $w(x)$. The class $\SHP$ is the class of counting problems naturally associated with decision problems in $\NP$. According to \cite{hevo95}  if ${\cal C}$ is a complexity class of decision problems, we define $\SHclass{\cal C}$ to be the class of all counting problems whose witness function is such that the size of every witness $y$ of $x$ is polynomially bounded in the size of $x$, and checking  whether $y\in w(x)$ is in ${\cal C}$. Thus, we have $\SHP=\SHclass{\P}$ and $\SHP\subseteq \SHcoNP$. Completeness of counting problems is usually proved by means of Turing reductions. A stronger notion is the parsimonious reduction where the exact number of solutions is conserved by the reduction function.

\paragraph{Propositional formulae}

  We assume familiarity with propositional logic. 
  The set of all propositional formulae is denoted by $\allFormulae$.
  A \emph{model} for a formula $\varphi$ is a truth assignment to the set of its variables that satisfies $\varphi$.
  Further we denote by $\varphi[\alpha/\beta]$ the formula obtaine from $\varphi$ by replacing all occurrences of $\alpha$ with $\beta$.
  For a given set $\Gamma$ of formulae, we write $\Vars{\Gamma}$ to denote the set of variables  occurring in $\Gamma$.
  We identify finite $\Gamma$ with the conjunction of all the formulae in $\Gamma$, $\bigwedge_{\varphi\in\Gamma}\varphi$. 
  For any formula $\varphi \in \allFormulae$, we write $\Gamma \models \varphi$ if $\Gamma$ entails $\varphi$, 
  \emph{i.e.}, if every model of $\Gamma$ also satisfies $\varphi$. 
  
  A \emph{literal} $l$ is a variable $x$ or its negation $\neg x$; $x$ is called the \emph{atom} of $l$ and is denoted by $\vert l \vert$. 
  Given a set of variables $V$, $\Lits{V}$ denotes the set of all literals formed upon the variables in $V\!$, \emph{i.e.},
  $\Lits{V}=V\cup \{\neg x\mid x\in V\}$.
  A \emph{clause} is a disjunction of literals 
  and a \emph{term} is a conjunction of literals.

\paragraph{Clones of Boolean Functions}

\begin{figure*}[t]
    \centering
    \hfill
    \includegraphics[height=0.6\textheight]{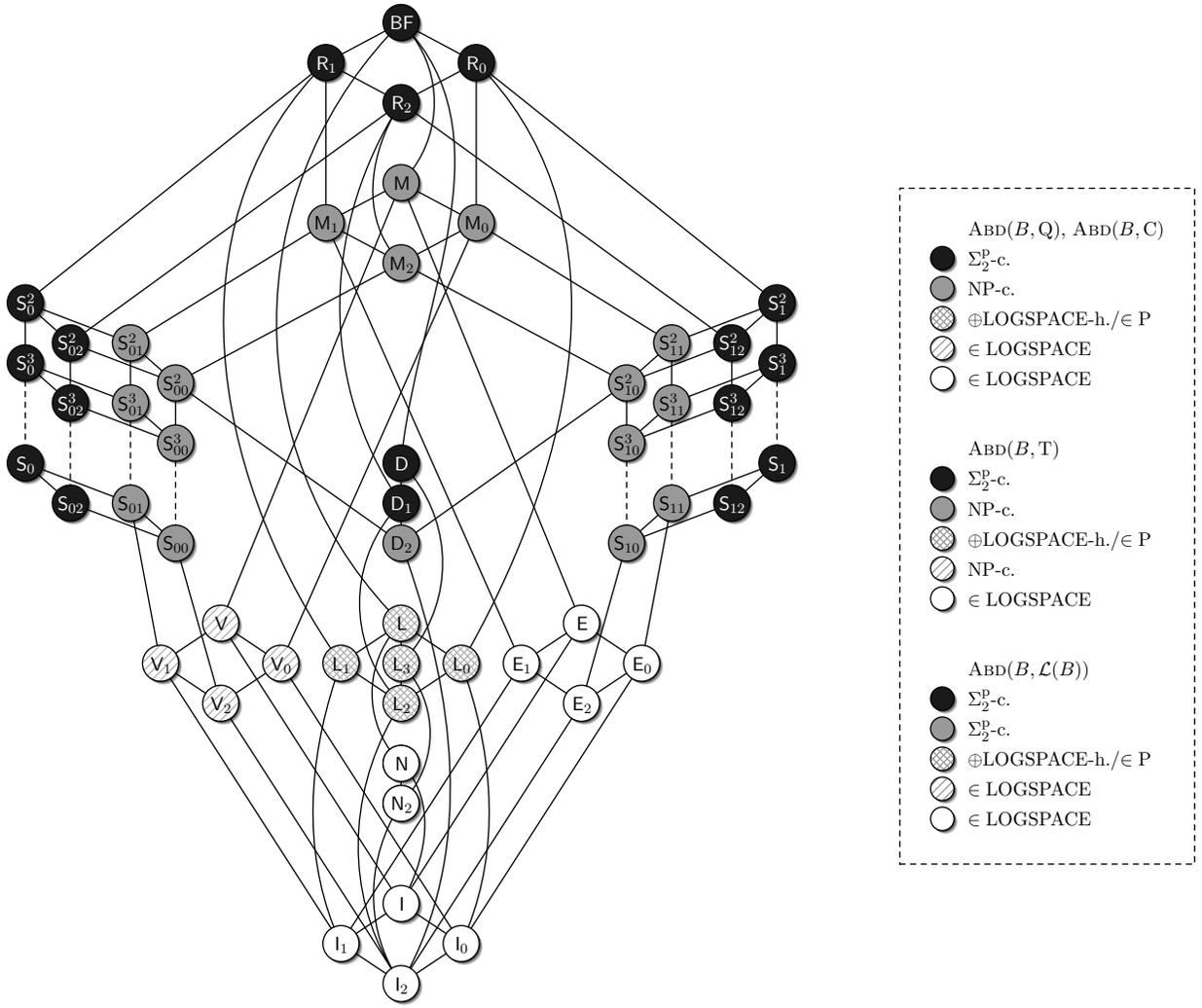}
    \caption{\label{fig:post's_lattice}Post's lattice showing the complexity of the abduction problem $\ABD(B,\manif)$ for all sets $B$
    	of Boolean functions and considered restrictions $\manif$ of the manifestations.}
  \end{figure*}

  A \emph{clone} is a set of Boolean functions that is closed under superposition, \emph{i.e.}, it  contains all projections (that is, the
  functions $f(a_1, \dots , a_n) = a_k$ for $1 \leq k \leq n$ and $n \in \N$) and is closed under arbitrary composition.
  Let $B$ be a
  finite set of Boolean functions. We denote by $[B]$ the smallest clone containing $B$ and call $B$ a \emph{base} for $[B]$.
  All closed classes of Boolean functions were identified by Post (\citeyear{pos41}). 
  Post also found a finite base for each of them and detected their
  inclusion structure, hence the name of \emph{Post's lattice} (see Figure~\ref{fig:post's_lattice}).

	In order to define the clones, we require the following notions, where $f$ is an $n$-ary Boolean function:
  \begin{itemize} \itemsep 0pt 
    \item $f$ is \emph{$c$-reproducing} if $f(c, \ldots , c) = c$, $c \in \{\false,\true\}$.
    \item $f$ is \emph{monotonic} if $a_1 \leq b_1, a_2 \leq b_2, \ldots , a_n \leq b_n$ implies $f(a_1, \ldots , a_n) \leq f(b_1, \ldots , b_n)$.
    \item $f$ is \emph{$c$-separating of degree $k$} if 
    for all $A \subseteq f^{-1}(c)$ of size $|A|=k$ 
    there exists an $i \in \{1, \ldots , n\}$ 
    such that $(a_1, \ldots , a_n) \in A$ implies $a_i = c$, $c \in \{\false,\true\}$.
    \item $f$ is \emph{$c$-separating} if 
    $f$ is $c$-separating of degree $|f^{-1}(c)|$.
    \item $f$ is \emph{self-dual} if $f \equiv \neg f(\neg x_1, \ldots , \neg x_n)$.
    \item $f$ is \emph{affine} if $f \equiv x_1 \xor \cdots \xor x_n \xor c$ with $c \in \{0, 1\}$.
  \end{itemize}
  
  A list of all clones with definitions and finite bases is given in Table~\ref{tab:clones} on page \pageref{tab:clones}, see also \emph{e.g.},
  \cite{bcrv03}. A propositional formula using only functions from $B$ as connectives is called a \emph{$B$-formula}.
  The set of all $B$-formulae is denoted by $\allFormulae(B)$.
  Let $f$ be an $n$-ary Boolean function. A $B$-formula $\varphi$ such that $\Vars{\varphi}= \{x_1,\ldots , x_n, y_1,\ldots , y_m\}$ is a
  \emph{$B$-representation} of $f$ if for all $a_1,\ldots , a_n, b_1,\dots , b_m \in \{\false,\true\}$ it holds that $f(a_1,\ldots , a_n)=1$
  if and only if every $\sigma\colon\Vars{\varphi}\longrightarrow\{\false,\true\}$ with $\sigma(x_i)=a_i$ and $\sigma(y_i)=b_i$ for all relevant $i$,
  satisfies $\varphi$. Such a $B$-representation exists for every $f\in[B]$. Yet, it may happen that the $B$-representation of some function uses
  some input variable more than once.
	\begin{example}
		Let $h(x,y)=x\wedge \neg y$. An $\{h\}$-representation of the function $x\land y$ is $h(x,h(x,y))$.
	\end{example}
	
  \begin{table*}[tb]
    \centering
    \small
    \begin{tabular}{p{1.2cm}ll}
    \toprule
      Name & Definition & Base \\
      \midrule[\heavyrulewidth]
      $\CloneBF$ & All Boolean functions & $\{x \land y, \neg x\}$ \\
      \sjhline
      $\CloneR_0$ & $\{f \mid f \text{ is $\false$-reproducing}\}$ & $\{x \land y, x \xor y\}$ \\
      \sjhline
      $\CloneR_1$ & $\{f \mid f \text{ is $\true$-reproducing}\}$ & $\{x \lor y, x \xor y \xor \true \}$ \\
      \sjhline
      $\CloneR_2$ & $\CloneR_0 \cap \CloneR_1$ & $\{\lor, x \land (y \xor z \xor \true) \}$ \\
      \sjhline
      $\CloneM$ & $\{f \mid f \text{ is monotonic}\}$ & $\{x \lor y, x \land y, \false, \true\}$ \\
      \sjhline
      $\CloneM_0$ & $\CloneM \cap \CloneR_0$ & $\{x \lor y, x \land y, \false\}$ \\
      \sjhline
      $\CloneM_1$ & $\CloneM \cap \CloneR_1$ & $\{x \lor y, x \land y, \true\}$ \\
      \sjhline
      $\CloneM_2$ & $\CloneM \cap \CloneR_2$ & $\{x \lor y, x \land y \}$ \\
      \sjhline

      $\CloneS^n_{0}$ & $\{f \mid f \text{ is $0$-separating of degree } n\}$ & $\{x \to y, \dual{h_n}\}$ \\
      \sjhline
      $\CloneS_0$ & $\{f \mid f \text{ is $0$-separating}\}$ & $\{x \to y\}$ \\
      \sjhline
      $\CloneS^n_{1}$ & $\{f \mid f \text{ is $1$-separating of degree } n\}$ & $\{x \wedge \neg y, h_n\}$ \\
      \sjhline
      $\CloneS_1$ & $\{f \mid f \text{ is $1$-separating}\}$ & $\{x \wedge \neg y\}$ \\
      \sjhline
      $\CloneS^n_{02}$ & $\CloneS^n_0 \cap \CloneR_2$ & $\{x \lor  (y  \land  \neg z), \dual{h_n}\}$ \\
      \sjhline
      $\CloneS_{02}$ & $\CloneS_0 \cap \CloneR_2$ & $\{x \lor  (y  \land  \neg z)\}$ \\
      \sjhline
      $\CloneS^n_{01}$ & $\CloneS^n_0 \cap \CloneM$ & $\{\dual{h_n},\true\}$ \\
      \sjhline
      $\CloneS_{01}$ & $\CloneS_0 \cap \CloneM$ & $\{x \lor  (y  \land z), \true\}$ \\
      \sjhline
      $\CloneS^n_{00}$ & $\CloneS^n_0 \cap \CloneR_2 \cap \CloneM$ & $\{x \lor  (y  \land  z), \dual{h_n}\}$ \\
      \sjhline
      $\CloneS_{00}$ & $\CloneS_0 \cap \CloneR_2 \cap \CloneM$ & $\{x \lor  (y  \land  z)\}$ \\
      \sjhline
      $\CloneS^n_{12}$ & $\CloneS^n_1 \cap \CloneR_2$ & $\{x \land  (y  \lor  \neg z), h_n\}$ \\
      \sjhline
      $\CloneS_{12}$ & $\CloneS_1 \cap \CloneR_2$ & $\{x \land  (y  \lor  \neg z)\}$ \\
      \sjhline
      $\CloneS^n_{11}$ & $\CloneS^n_1 \cap \CloneM$ & $\{h_n, \false\}$ \\
      \sjhline
      $\CloneS_{11}$ & $\CloneS_1 \cap \CloneM$ & $\{x \land  (y  \lor  z), \false\}$ \\
      \sjhline
      $\CloneS^n_{10}$ & $\CloneS^n_1 \cap \CloneR_2 \cap \CloneM$ & $\{x \land  (y  \lor  z), h_n\}$ \\
      \sjhline
      $\CloneS_{10}$ & $\CloneS_1 \cap \CloneR_2 \cap \CloneM$ & $\{x \land  (y  \lor  z)\}$ \\
      \sjhline

      $\CloneD$ & $\{f \mid f \text{ is self-dual}\}$ & $ \{(x \land \neg y) \lor (x \land \neg z) \lor (\neg y \land  \neg z)\}$ \\
      \sjhline
      $\CloneD_1$ & $\CloneD \cap \CloneR_2$ & $ \{(x \land y) \lor (x \land \neg z) \lor (y \land \neg z)\}$ \\
      \sjhline
      $\CloneD_2$ & $\CloneD \cap \CloneM$ & $ \{(x \land y) \lor (x \land z) \lor (y \land z)\}$ \\
      \sjhline

      $\CloneL$ & $\{f \mid f \text{ is affine}\}$ & $\{ x \xor y,\true\}$ \\
      \sjhline
      $\CloneL_0$ & $\CloneL \cap \CloneR_\false$ & $\{x \xor y \}$ \\
      \sjhline
      $\CloneL_1$ & $\CloneL \cap \CloneR_\true$ & $\{x \xor y  \xor \true \}$ \\
      \sjhline
      $\CloneL_2$ & $\CloneL \cap \CloneR_2$ & $\{x \xor y \xor z\}$ \\
      \sjhline
      $\CloneL_3$ & $\CloneL \cap \CloneD$ & $\{x  \xor  y  \xor  z  \xor  \true\}$ \\
      \sjhline

      $\CloneV$ & $\{f \mid  f $ is a disjunction of variables or constants$\}$ & $\{ x \lor y, \false,\true \}$ \\
      \sjhline
      $\CloneV_0$ & $\CloneV \cap \CloneR_0$ & $\{ x \lor y, \false\}$ \\
      \sjhline
      $\CloneV_1$ & $\CloneV \cap \CloneR_1$ & $\{ x \lor y, \true\}$ \\
      \sjhline
      $\CloneV_2$ & $\CloneV \cap \CloneR_2$ & $\{ x \lor y\}$ \\
      \sjhline

      $\CloneE$ & $\{f \mid  f $ is a conjunction of variables or constants$\}$ & $\{ x \land y, \false, \true \}$ \\
      \sjhline
      $\CloneE_0$ & $\CloneE \cap \CloneR_0$ & $\{ x \land y, \false\}$ \\
      \sjhline
      $\CloneE_1$ & $\CloneE \cap \CloneR_1$ & $\{ x \land y, \true\}$ \\
      \sjhline
      $\CloneE_2$ & $\CloneE \cap \CloneR_2$ & $\{ x \land y\}$ \\
      \sjhline

      $\CloneN$ & $\{f \mid f $ depends on at most one variable$\}$ & $\{ \neg x,\false,\true\}$ \\
      \sjhline
      $\CloneN_2$ & $\CloneN \cap \CloneR_2$ & $\{ \neg x\}$ \\
      \sjhline

      $\CloneI$ & $\{f \mid f \text{ is a projection or a constant}\}$ & $\{\id, \false,\true\}$ \\
      \sjhline
      $\CloneI_0$ & $\CloneI \cap \CloneR_0$ & $\{\id, \false\}$ \\
      \sjhline
      $\CloneI_1$ & $\CloneI \cap \CloneR_1$ & $\{\id, \true\}$ \\
      \sjhline
      $\CloneI_2$ & $\CloneI \cap \CloneR_2$ & $\{\id\}$ \\
      \bottomrule
    \end{tabular}
    \smallskip
    \caption{\label{tab:clones}%
      The list of all Boolean clones with definitions and bases, where $h_n := \bigvee^{n+1}_{i=1}\bigwedge^{n+1}_{j=1,j\neq i} x_j$ and
      $\dual{f}(a_1, \dots , a_n) = \neg f(\neg a_1 \dots , \neg a_n)$.
    }
  \end{table*}

Observe that if $B_1$ and $B_2$ are two sets of Boolean functions such that $B_1\subseteq [B_2]$, then every function of $B_1$ can be expressed by a $B_2$-formula, its so-called \emph{$B_2$-representation}.

\section{The Abduction Problem} \label{sect:abduction-def}

  Let $B$ be a finite set of Boolean functions. We are interested in a propositional abduction problem
	parameterized by the set $B$ of allowed connectives. 
  We define the \emph{abduction problem for $B$-formulae} as
  \problemdef{$\ABD(B)$}
      {$\calP=(\Gamma,A,\varphi)$, where
      \begin{itemize}
        \item $\Gamma$ is a set of $B$-formulae,  $\Gamma\subseteq\allFormulae(B)$,
        \item $A$ is a set of variables, $A\subseteq \Vars{\Gamma}$,
        \item $\varphi$ is a formula, $\varphi \in \allFormulae$ with $\Vars{\varphi} \subseteq \Vars{\Gamma}\setminus A.$
      \end{itemize}
      }
      {Is there a set $E \subseteq \Lits{A}$ such that $\Gamma \land E$ is satisfiable and $\Gamma \land E \models \varphi$ (or equivalently $\Gamma \land E \land \neg\varphi$ is unsatisfiable)?}
  The set $\Gamma$ represents the \emph{knowledge base}. The set $A$ is called the set of \emph{hypotheses} and $\varphi$ is called \emph{manifestation}
  or \emph{query}.
  Furthermore, if such a set $E$ exists, it is called an \emph{explanation} or a \emph{solution} of the abduction problem. 
  It is called  a \emph{full explanation} if $\Vars{E}=A$. Observe that every explanation can be extended to a full one.
  
  We will consider several restrictions on the manifestations of this problem.
  To indicate these restrictions, we introduce a second argument $\manif$: 
  in the abduction problem $\ABD(B,\manif)$, $\varphi$ is required to be a single literal if $\manif=\MQ$, 
  a clause if $\manif=\MC$, a term if $\manif=\MT$, and a $B$-formula if $\manif=\MF$.
   
  Let us start with a lemma that makes clear the role of the two constants $0$ and $1$ in our problem.

  \begin{lemma} \label{lem:constants_sometimes_available}
  Let $B$ be a finite set of Boolean functions
  \begin{enumerate}
  	\item If $\manif \in \{\MQ, \MC, \MT, \MF\}$, then 
  				$$\ABD (B,\manif) \equivlogm \ABD (B \cup \{\true\},\manif)$$
  	\item If $\manif \in \{\MQ, \MC, \MT\}$ and $\lor\in [B]$, then 
  				$$\ABD (B,\manif) \equivlogm \ABD (B \cup \{\false\},\manif)$$
  \end{enumerate}
  \end{lemma}

  \begin{proof}
  	To reduce $\ABD (B \cup \{\true\},\manif)$ to $\ABD (B,\manif)$ we transform any instance of the first problem in replacing every occurrence of
  	$\true$ by a fresh variable $t$ and adding the unit clause $(t)$ to the knowledge base.
  	To prove $\ABD (B \cup \{\false\},\manif)\equivlogm\ABD (B,\manif) $, let $\calP=(\Gamma,A,\psi)$  be an instance of the first problem and $f$ be a
  	fresh variable. If $\manif \in \{\MQ,\MC,\MT\}$, then we can suppose w.l.o.g.\ that $\psi$ does not contain $\false$. We map $\calP$ to
  	$\calP'=(\Gamma', A\cup\{f\}, \psi)$, where $\Gamma'$ is the $B$-representation of $\{\varphi[\false/f]\lor f\mid \varphi\in\Gamma\}$.
 	\end{proof}

\section{The Complexity of $\ABD(B,\MQ)$} \label{sect:abduction-literals}

  \begin{theorem}\label{thm:main}
    Let $B$ be a finite set of Boolean functions. Then, the abduction problem for propositional $B$-formulae, $\ABD(B,\MQ)$,  is
    \begin{enumerate}
      \item $\SigPtwo$-complete if $\CloneS_{02}\subseteq [B]$ or $\CloneS_{12}\subseteq [B]$ or  $\CloneD_{1}\subseteq [B]$,
      \item $\NP$-complete if  $\CloneS_{00}\subseteq [B]\subseteq \CloneM$ or $\CloneS_{10}\subseteq [B]\subseteq \CloneM$ or
      			$\CloneD_{2}\subseteq [B]  \subseteq \CloneM$,
      \item in $\P$ and $\ParityL$-hard if $\CloneL_{2}\subseteq [B]\subseteq \CloneL$, and
      \item in $\L$ in all other cases.
    \end{enumerate}
  \end{theorem}

	\begin{remark}\label{rem:recognize_the_clone}
		For such a classification a natural question is: given $B$, how hard is it to determine the complexity of $\ABD(B,\MQ)$? Solving this
		task requires checking whether certain clones are included in $[B]$ (for lower bounds) and whether $B$ itself is included in certain clones
		(for upper bounds). As shown in \cite{vol09}, the complexity of checking whether certain Boolean functions are included in a clone depends on the
		representation of the Boolean functions. If all functions are
		given by their truth table then the problem is in quasi-polynomial-size $\AC{0}$, while if the input functions are given in a compact way,
		\emph{i.e.}, by circuits, then the above problem becomes $\coNP$-complete.
	\end{remark}

  We split the proof of Theorem~\ref{thm:main} into several propositions.

  \begin{proposition}\label{prop:abd(q)-logspace}
    Let $B$ be a finite set of Boolean functions such that $[B] \subseteq \CloneE$ or $[B] \subseteq \CloneN$ or $[B] \subseteq \CloneV$.
    Then $\ABD(B,\MQ) \in \L$.
  \end{proposition}
  
  \begin{proof}
    Let $\calP=(\Gamma,A,q)$ be an instance of $\ABD(B,\MQ)$.
    
    For $[B] = \CloneN$ or $\CloneE$, $\Gamma$ is equivalent to a set of literals,
    hence $\calP$ has the empty set as a solution if $\calP$ possesses a solution at all.
    Finally notice that satisfiability of a set of $\CloneN$-formulae can be tested in logarithmic space \cite{sch05}.
     
  	For $[B] = \CloneV$ each formula $\varphi \in \Gamma$ is equivalent to either a constant or disjunction.
  	It holds that  $(\Gamma,A,q)$ has a solution if and only if  $\Gamma$ contains a formula
		$\varphi \equiv q \lor x_1 \lor \cdots \lor x_k$ such that $X:=\{x_1, \dots x_k\} \subseteq A$, and $\Gamma[X/\false]$ is satisfiable. This can be
		tested in logarithmic space, as substitution of symbols and  evaluation of $\CloneV$-formulae can all be performed
  	in logarithmic space.
  
  \end{proof}

  \begin{proposition}\label{prop:abd(q)-l-p}
    Let $B$ be a finite set of Boolean functions such that $\CloneL_2 \subseteq [B] \subseteq \CloneL$. 
    Then $\ABD(B,\MQ)$ is $\ParityL$-hard and contained in $\P$.
  \end{proposition}
  \begin{proof}
	  In this case, deciding whether an instance of $\ABD(B,\MQ)$ has a solution logspace reduces to the problem of deciding whether a propositional
	  abduction problem in which the knowledge base is a set of linear equations has a solution. This has been shown to be decidable in polynomial
	  time in \cite{zanuttini03}.
  

    As for the $\ParityL$-hardness, let $B$ be such that $[B]=\CloneL_2$.
    Consider the $\ParityL$-complete problem to determine whether a system of linear equations $S$ over $GF(2)$ has a solution \cite{budaheme92}.
    Note that  $\ParityL$ is closed under complement, so deciding whether such a system has no solution is also  $\ParityL$-complete.
    Let $S=\{s_1,\ldots,s_m\}$ be such a system of linear equations over variables $\{x_1,\ldots,x_n\}$.
    Then, for all $1 \leq i \leq m$, the equation $s_i$ is of the form $ x_{i_1} + \cdots + x_{i_{n_i}} = c_i  \pmod 2$ with $c_i \in \{0,1\}$
    and $i_1,\ldots,i_{n_i} \in \{1,\ldots,n\}$.
    We map $S$ to a set of affine formulae $\Gamma=\{\varphi_1,\ldots,\varphi_m\}$ over variables $\{x_1,\ldots,x_n,q\}$ via
    \[
      \begin{array}{lll}
        \varphi_i &:= x_{i_1} \xor \cdots \xor x_{i_{n_i}} \xor \true & \text{ if $c_i = 0$, and} \\
        \varphi_i &:= x_{i_1} \xor \cdots \xor x_{i_{n_i}}            & \text{ if $c_i = 1$}.
      \end{array}
    \]
    Now define 
    \begin{align*}
      \Gamma' := \,
            & \{\varphi_i \xor q \mid \varphi_i \in \Gamma  \text{ such that }\varphi_i(1,\ldots, 1)=0\} \\
      \cup \; & \{\varphi_i \mid \varphi_i \in \Gamma \text{ such that{} }\varphi_i(1,\ldots, 1)=1\}.
    \end{align*}
    $\Gamma'$ is obviously satisfied by the assignment mapping all propositions to $\true$.
    It furthermore holds that $S$ has no solution if and only if $\Gamma' \land \neg q$ is unsatisfiable. Hence, we obtain that $S$ has no solution if
    and only if the propositional abduction problem $(\Gamma',\emptyset,q)$ has an explanation. 

		It remains to transform $\Gamma'$ into a set of $B$-formulae in logarithmic space. Since $[B]=\CloneL_2$, we have $x \xor y \xor z\in[B]$. 
		We insert parentheses in every formula $\varphi$  of $\Gamma'$ in such a way that we get a ternary $\xor$-tree of logarithmic depth
		whose leaves are either a proposition or the constant 1. Then we replace every node $\xor$ by its equivalent $B$-formula. Thus we get a
		$(B\cup\{1\})$-formula of size  polynomial in the size of the original one. Lemma~\ref{lem:constants_sometimes_available} allows to conclude.

		Note that the $B$-formulae replacing the connectives might use some input variable more than once. Therefore, 
		the logarithmic depth tree is built in order to avoid an exponential explosion of the formula size during the replacement.
  \end{proof}

		Observe that the abduction problem for $B$-formulae is self-reducible for the above cases, \emph{i.e.}, for
		$[B] \subseteq \CloneL$, $[B] \subseteq \CloneE$ and $[B] \subseteq \CloneV$.
    Roughly speaking this means, given an instance $\calP$ and a literal $l$,
		we can compute efficiently an instance $\calP'$ such that
		the question whether there exists an explanation $E$ with $l \in E$ 
		reduces to the question whether $\calP'$ admits solutions. 
		It is well-known that for self-reducible problems whose decision problem is in $\P$, the
		lexicographically first solution can be computed in $\FP$. It is an easy exercise to extend this algorithm to enumerate all
		solutions in lexicographical order with polynomial delay and polynomial space. Thus, the explanations of $\ABD(B,\MQ)$ can be enumerated with
		polynomial delay and polynomial space if $[B] \subseteq \CloneL$ or	$[B] \subseteq \CloneE$ or $[B] \subseteq \CloneV$,
		according to Proposition~\ref{prop:abd(q)-logspace} and \ref{prop:abd(q)-l-p}.


  \begin{proposition}\label{prop:abd(q)-m-np}
    Let $B$ be a finite set of Boolean functions such that $\CloneS_{00} \subseteq [B] \subseteq \CloneM$ or
    $\CloneS_{10} \subseteq [B] \subseteq \CloneM$ or $\CloneD_{2} \subseteq [B] \subseteq \CloneM$. Then $\ABD(B,\MQ)$ is $\NP$-complete.
  \end{proposition}
  
  \begin{proof}
    We first show that $\ABD(B,\MQ)$ is efficiently verifiable. 
    Let $\calP=(\Gamma,A,q)$ be an $\ABD(B,\MQ)$-instance and $E \subseteq \Lits{A}$ be a candidate for an explanation. 
    Define $\Gamma'$ as the set of formulae obtained from $\Gamma$ by replacing each occurrence of the proposition $x$ with $\false$ if $\neg x \in E$,
    and each occurrence of the proposition $x$ with $\true$ if $x \in E$.
    It holds that $E$ is a solution for $\calP$  if $\Gamma'$ is satisfiable and $\Gamma'[q/\false]$ is not. 
    These tests can be performed in polynomial time, because $\Gamma'$ is a set of monotonic formulae \cite{lew79}.
    Hence, $\ABD(B,\MQ) \in \NP$.
    
    Next we give a reduction from the $\NP$-complete problem $\TWOINTHREESAT$, \emph{i.e.}, the problem to decide whether there exists an assignment that
    satisfies exactly two propositions in each clause of a given formula in conjunctive normal form with exactly three positive propositions
    per clause, see \cite{sch78}.
    Let $\varphi := \bigwedge_{i \in I} c_i$ with $c_i=x_{i1} \lor x_{i2} \lor x_{i3}$, $i \in I$, be the given formula.
    We map $\varphi$ to the following instance $\calP=(\Gamma,A,q)$.
    Let $q$, $q_i$, $i \in I$, be fresh, pairwise distinct propositions and let $A:= \Vars{\varphi} \cup \{q_i\mid i \in I\}$.
    We define $\Gamma$ as 
    \begin{align}
      \Gamma :=\,& \label{eq:q-abd-s01-1}
      \{c_i | i \in I \} \\
       \cup\; & \label{eq:q-abd-s01-2}
      \{ x_{i1} \!\lor\! x_{i2} \!\lor\! q_i, x_{i1} \!\lor\! x_{i3} \!\lor\! q_i, x_{i2} \!\lor\! x_{i3} \!\lor\! q_i | i \in I \} \\  
       \cup\; & \textstyle \label{eq:q-abd-s01-3}
      \{\bigvee_{i \in I} \bigwedge_{j=1}^3 x_{ij} \lor \bigvee_{i \in I} q_i \lor q\} 
    \end{align}
    We show that there is an assignment that sets to true exactly two propositions in each clause of $\varphi$ if and only if $\calP$ has a solution. 
    First, suppose that there exists an assignment $\sigma$ such that for all $i \in I$, there is a permutation $\pi_i$ of $\{1,2,3\}$ such that\
    $\sigma(x_{i\pi_i(1)})=\false$ and $\sigma(x_{i\pi_i(2)})=\sigma(x_{i\pi_i(3)})=\true$. 
    Thus \eqref{eq:q-abd-s01-1} and \eqref{eq:q-abd-s01-2} are satisfied,
    and \eqref{eq:q-abd-s01-3} is equivalent to $\bigvee_{i \in I} q_i \lor q$.
    From this, it is readily observed that 
    $\{ \neg x \mid \sigma(x)=\false\} \cup \{\neg q_i \mid i \in I\}$ is a solution to $\calP$.

    Conversely, suppose that $\calP$ has an explanation $E$ that is w.l.o.g.\ full.
    Then $\Gamma \land E$ is satisfiable and $\Gamma \land E \models q$. 
    Let  $\sigma \colon \Vars{\Gamma}  \to \{\false,\true\}$ be an assignment that satisfies $\Gamma \land E$. Then, for any $x \in A$,
    $\sigma(x)=\false$ if $\neg x \in E$, and $\sigma(x)=\true$ otherwise. Since $\Gamma \land E$ entails $q$
    and as the only occurrence of $q$ is in \eqref{eq:q-abd-s01-3}, 
    we obtain  that
		$\sigma$ sets to $\false$ each $q_i$ and at least one proposition in each clause of $\varphi$.
		Consequently, from \eqref{eq:q-abd-s01-2} follows that $\sigma$ sets to $\true$  at least two propositions in each clause of $\varphi$.
    Therefore, $\sigma$ sets to $\true$ exactly two propositions in each clause of $\varphi$.
    

		It remains to show that $\calP$ can be transformed into an $\ABD(B,\MQ)$-instance for all considered $B$.
		Observe that $\lor \in [B\cup\{1\}]$ and
		$[\CloneS_{00}\cup\{\false, \true\}] = [\CloneD_{2}\cup\{\false, \true\}] = [\CloneS_{10}\cup\{\false, \true\}] = \CloneM$. 
		Therefore due to Lemma~\ref{lem:constants_sometimes_available} it suffices to consider the case $[B]=\CloneM$.
		Using the associativity of $\lor$ rewrite \eqref{eq:q-abd-s01-3} as an $\lor$-tree of logarithmic depth and replace all the connectives in
		$\Gamma$ by their B-representation ($\lor,\land \in [B]$).
  \end{proof}

  \begin{proposition}\label{prop:abd(q)-bf-sigP2}
    Let $B$ be a finite set of Boolean functions such that $\CloneS_{02} \subseteq [B]$ or $\CloneS_{12} \subseteq [B]$ or $\CloneD_{1} \subseteq [B]$.
    Then $\ABD(B,\MQ)$ is $\SigPtwo$-complete.
  \end{proposition}
  
  \begin{proof}
    Membership in $\SigPtwo$ is easily seen to hold: given an instance $(\Gamma,A,q)$,
   	guess an explanation $E$ and subsequently verify that $\Gamma \land E$ is satisfiable and $\Gamma \land E \land \neg q$ is not.
   	
    Observe that $\lor \in [B\cup\{\true\}]$. By virtue of Lemma~\ref{lem:constants_sometimes_available} and the fact that
    $[\CloneS_{02} \cup \{\false,\true\}] = [\CloneS_{12} \cup \{\false,\true\}] = [\CloneD_{1} \cup \{\false,\true\}] = \CloneBF$,  
    it suffices to consider the case $[B] = \CloneBF$.
    In \cite{eiter-gottlob:1995} it has been shown that the propositional abduction problem
    remains $\SigPtwo$-complete when the knowledge base $\Gamma$ is a set of CNF-formulae.
		From such an instance $(\Gamma, A, q)$ we build an instance of $\ABD(B,\MQ)$ by rewriting first each formula as a tree of logarithmic depth
		and then replacing all the connectives $\land$,$\lor$ and $\neg$ by their $B$-representation, thus concluding the proof.
	\end{proof}

\section{Complexity of the Variants} \label{sect:abduction-variants}

  We now turn to the study of the complexity of some variants of the abduction problem.
  It is obvious that  $\ABD(B,\MQ)\leqlogm \ABD(B,\MC)$ and that $\ABD(B,\MQ)\leqlogm \ABD(B,\MT)$. Therefore, all hardness results still hold for the
  variants $\ABD(B,\MC)$ and $\ABD(B,\MT)$.  Also, it can be easily checked that the hardness results in the previous sections still hold 
  when the query is required to be a positive literal.  For this reason the hardness results also carry over to the variant $\ABD(B,\MF)$.

  It is an easy exercise to prove that all  algorithms that have been developed for a single query can be naturally extended to clauses. Therefore, the
  complexity classification for the  problem $\ABD(B,\MC)$ is exactly the same as for $\ABD(B,\MQ)$.
  
  \begin{theorem}\label{thm:abd(c)-all}
    Let $B$ be a finite set of Boolean functions. Then, the abduction problem for propositional $B$-formulae, $\ABD(B,\MC)$,  is
    \begin{enumerate}
      \item $\SigPtwo$-complete if $\CloneS_{02}\subseteq [B]$ or $\CloneS_{12}\subseteq [B]$ or  $\CloneD_{1}\subseteq [B]$,
      \item $\NP$-complete if  $\CloneS_{00}\subseteq [B]\subseteq \CloneM$ or $\CloneS_{10}\subseteq [B]\subseteq \CloneM$ or
      			$\CloneD_{2}\subseteq [B]  \subseteq \CloneM$,
      \item in $\P$ and $\ParityL$-hard if $\CloneL_{2}\subseteq [B]\subseteq \CloneL$, and
      \item in $\L$ in all other cases.
    \end{enumerate}
  \end{theorem}

  More interestingly, we will prove in the next section that allowing terms as  manifestations increases the complexity for the clones $\CloneV$ (from
  membership in $\L$ to $\NP$-completeness), while allowing  $B$-formulae as manifestations makes the classification dichotomous,
  $\P$/$\SigPtwo$-complete, thus skipping the intermediate $\NP$ level.

  \subsection{The Complexity of $\ABD(B,\MT)$}

  \begin{proposition}\label{prop:abd(t)-v-np}
    Let $B$ be a finite set of Boolean functions such that $\CloneV_2 \subseteq [B] \subseteq \CloneV$. Then $\ABD(B,\MT)$ is $\NP$-complete.
  \end{proposition}
  
  \begin{proof}
    Let $B$ be a finite set of Boolean functions such that\ $\CloneV_2 \subseteq [B] \subseteq \CloneV$ and let $\calP=(\Gamma,A,t)$ be an instance of
    $\ABD(B,\MT)$. Hence, $\Gamma$ is a set of $B$-formulae and $t$ is a term, $t=\bigwedge_{i=1}^n l_i$. Observe that $E$ is a solution for $\calP$
    if $\Gamma\land E$ is satisfiable and for every $i=1,\ldots, n$, $\Gamma\land E\land \neg l_i$ is not. Given a set $E\subseteq\Lits{A}$,  these
    verifications, which  require substitution of symbols and evaluation of an $\lor$-formula,  can be performed in polynomial time, thus proving
    membership in $\NP$.

    To prove $\NP$-hardness, we give a reduction from $\ThreeSAT$. 
    Let $\varphi$ be a 3-CNF-formula, 
    $\varphi := \bigwedge_{i \in I} c_i$.
    Let $x_{1}, \ldots, x_{n}$ enumerate the variables occurring in $\varphi$. 
    Let $x'_{1},\ldots, x'_{n}$ and $q_{1}, \ldots, q_{n}$ be fresh, pairwise distinct variables.
    We map $\varphi$ to $\calP=(\Gamma,A,t)$, where
    \begin{align*}
      \Gamma :=\,& \{ c_i[\neg x_1/x'_1,\ldots, \neg x_n/x'_n]\mid i\in I\}\\
      \cup\;& \{ x_i \lor x'_i,   x_i \lor q_i,  x'_i \lor q_i \mid 1 \leq i \leq n\}, \\
      A:=\,& \{x_1,\ldots,x_n,x'_1,\ldots,x'_n\}, \\
      t :=\,& q_1\land \cdots \land q_n.
    \end{align*}

    We show that $\varphi$ is satisfiable if and only if $\calP$ has a solution.
    First assume that $\varphi$ is satisfied by the assignment $\sigma\colon \{x_{1}, \ldots, x_{n}\} \to \{\false,\true\}$. 
    Define $E:= \{ \neg x_i \mid \sigma(x_i)=\false \} \cup \{ \neg x'_i \mid \sigma(x_i)=\true \}$ and 
    $\hat\sigma$ as the extension of $\sigma$ mapping $\hat\sigma(x'_i)=\neg \sigma(x_i)$ and $\hat\sigma(q_i)=\true$ for all $1 \leq i \leq n$. 
    Obviously, $\hat\sigma \models \Gamma \land E$. 
    Furthermore, $\Gamma \land E \models q_i$ for all $1 \leq i \leq n$, 
   	because any satisfying assignment of $\Gamma \wedge E$ sets to $\false$ either $x_i$ or $x_i'$ and thus $\{  x_i \lor q_i,  x'_i \lor q_i\}
   	\models q_i$. Hence $E$ is an explanation for $\calP$.
    
    Conversely, suppose that $\calP$ has a full explanation $E$. The facts that  $\Gamma\land E\models q_1\land \cdots \land q_n $  and that each
    $q_i$ occurs only in the clauses  $ x_i \lor q_i,  x'_i \lor q_i$ enforce that, for every $i$, $E$  contains $\neg x_i$ or $\neg x'_i$. Because of
    the clause $x_i \lor x'_i$, it cannot contain both. Therefore in $E$ the value of $x'_i$ is determined by the value of $x_i$ and is its dual. From
    this it is easy to conclude that the assignment $\sigma\colon \{x_{1}, \ldots, x_{n}\} \to \{\false,\true\}$ defined by $\sigma(x_i)=\false$ if
    $\neg x_i\in E$, and $\true$ otherwise, satisfies $\varphi$.
    Finally $\calP$ can be transformed into an $\ABD(B,\MT)$-instance, because every formula in $\Gamma$ is the disjunction of at most three variables
    and $\lor\in [B]$.
  \end{proof}

    \begin{theorem}\label{thm:main_variant_T}
      Let $B$
      be a finite set of Boolean functions. Then, the abduction problem for propositional $B$-formulae, $\ABD(B,\MT)$, is
      \begin{enumerate}
        \item $\SigPtwo$-complete if $\CloneS_{02}\subseteq [B]$ or $\CloneS_{12}\subseteq [B]$ or  $\CloneD_{1}\subseteq [B]$,
        \item $\NP$-complete if  $\CloneV_2\subseteq [B]\subseteq \CloneM$ or $\CloneS_{10}\subseteq [B]\subseteq \CloneM$ or
        			$\CloneD_{2}\subseteq [B] \subseteq \CloneM$,
        \item in $\P$ and $\ParityL$-hard if $\CloneL_{2}\subseteq [B]\subseteq \CloneL$, and
        \item in $\L$ in all  other cases.
      \end{enumerate}
    \end{theorem}

  \subsection{The Complexity of $\ABD(B,\MF)$}
    \begin{proposition}
      Let $B$ be a finite set of Boolean functions such that $\CloneS_{00} \subseteq [B]$ or $\CloneS_{10} \subseteq [B]$ or $\CloneD_2 \subseteq [B]$.
      Then $\ABD(B,\MF)$ is $\SigPtwo$-complete.
    \end{proposition}
    
    \begin{proof}
      We prove $\SigPtwo$-hardness by giving a reduction from the $\SigPtwo$-hard problem $\QSAT_2$ \cite{wra77}. Let an instance of $\QSAT_2$ be given
      by a closed formula $\chi:=\exists x_1 \cdots \exists x_n \forall y_1 \cdots \forall y_m \varphi$ with $\varphi$ a 3-DNF-formula.
      First observe that $\exists x_1 \cdots \exists x_n \forall y_1 \cdots \forall y_m \varphi$ is true if and only if there exists a consistent set
      $X \subseteq \Lits{\{x_1,\ldots,x_n\}}$ such that $X \cap \{x_i,\neg x_i\} \neq \emptyset$, for all $1 \leq i \leq n$, and $\neg X \lor \varphi$
      is (universally) valid (or equivalently $\neg\varphi \land X$ is unsatisfiable). 

      Denote by $\overline{\varphi}$ the negation normal form of $\neg \varphi$ and
      let $\overline{\varphi}'$   be obtained from $\overline{\varphi}$ by replacing 
      all occurrences of $\neg x_i$ with a fresh proposition $x_i'$, $1\leq i \leq n$, and
      all occurrences of $\neg y_i$ with a fresh proposition $y_i'$, $1\leq i \leq m$.
      That is, $\overline{\varphi}' \equiv \overline{\varphi}[\neg x_1/x'_1,\ldots,\neg x_n/x'_n, \neg y_1/y'_1,\ldots,\neg y_m/y'_m]$. Thus
      $\overline{\varphi}'=\bigwedge _{i\in I} c'_i$ where every $c'_i$ is a disjunction of  three propositions. To $\chi$ we associate the
      propositional abduction problem $  \calP=(\Gamma,A,\psi)$ defined as follows:
      \begin{align*}
      \Gamma:=\,& \{c'_i \lor q\mid   i \in I\} \\ 
         \cup\;& \{ x_i \lor x'_i \mid 1 \leq i \leq n\} \cup \{ y_i \lor y'_i \mid 1 \leq i \leq m\} \\
         \cup\;& \{f_i \lor x_i, t_i \lor x'_i, f_i \lor t_i \mid 1 \leq i \leq n \}, \\
          A :=\,& \{t_i,f_i \mid 1 \leq i \leq n \},\\
        \psi :=\,&\textstyle q \lor \bigvee_{1 \leq i \leq n} (x_i \land x'_i) \lor \bigvee_{1 \leq i \leq m} (y_i \land y'_i).
      \end{align*}
     
      Suppose that $\chi$ is true.
      Then there exists an assignment $\sigma\colon\{x_1,\ldots,x_n\} \to \{\false,\true\}$ such that no extension $\sigma'\colon\{x_1,\ldots,x_n\} \cup
      \{y_1,\ldots,y_m\} \to \{\false,\true\}$ of $\sigma$ satisfies $\neg \varphi$.
      Define $X$ as the set of literals over $\{x_1,\ldots,x_n\}$ set to $\true$ by $\sigma$.
      Defining $E:= \{ \neg f_i, t_i \mid x_i \in X \} \cup \{ \neg t_i, f_i \mid \neg x_i \in X \}$, we obtain with abuse of notation
      \begin{align*}
                & \Gamma \land E \land \neg \psi \\
      \equiv \; &\textstyle \bigwedge_{i \in I} c'_i \land \bigwedge_{1 \leq i \leq n} (x_i \xor x'_i)\land \bigwedge_{1 \leq i \leq m}
      					(y_i \xor y'_i)  \land  \\ 
                &\textstyle \bigwedge_{1 \leq i \leq n, \sigma(x_i)=1} x_i \land \bigwedge_{1 \leq i \leq n, \sigma(x_i)=0} x'_i \\
      \equiv \; & \neg\varphi \land X,
      \end{align*}
      
      which is unsatisfiable by assumption.
      As $\Gamma \land E$ is satisfied by any assignment setting in addition all $x_i, x'_i$, $1 \leq i \leq n$, and all $y_j, y'_j$, $1 \leq i \leq m$, 
      to $\true$, we have proved that $E$ is an explanation for $\calP$.

      Conversely, suppose that $\calP$ has an explanation $E$. 
      Due to the clause $(f_i\lor t_i)$ in $\Gamma$,
      we also may assume that $|E \cap \{\neg t_i, \neg f_i\}| \leq 1$ for all $1 \leq i \leq n$. 

      Setting $X := \{x_i \mid \neg f_i \in E\} \cup \{\neg x_i \mid \neg t_i \in E\}$ we now obtain
      $\bigwedge_{1 \leq i \leq n} \big((f_i \lor x_i) \land (t_i \lor \neg x_i)\land (f_i\lor t_i)\big) \land E\equiv X$ and 
      $\Gamma \land E \land \neg \psi \equiv \neg\varphi \land X$ as above.
      Hence, $\neg\varphi \land X$ is unsatisfiable, which implies the existence of an assignment 
      $\sigma\colon\{x_1,\ldots,x_n\} \to \{\false,\true\}$ such that no extension 
      $\sigma'\colon\{x_1,\ldots,x_n\} \cup \{y_1,\ldots,y_m\} \to \{\false,\true\}$ of $\sigma$ satisfies $\neg \varphi$.
      Therefore, we have proved that $\chi$ is true if and only if $\calP$ has an explanation. 

      It remains to show that $\calP$ can be transformed into an $\ABD(B,\MF)$-instance for any relevant $B$.
      Since $[\CloneS_{00} \cup \{\true\}] = \CloneS_{01}$, $[\CloneS_{10} \cup \{\true\}] = \CloneM_1$, $[\CloneD_2 \cup \{\true\}] = \CloneS^2_{01}$ 
      and $\CloneS_{01} \subseteq \CloneS^2_{01} \subseteq\CloneM_1$, by Lemma~\ref{lem:constants_sometimes_available} it suffices to consider
      the case $[B] = \CloneS_{01}$. Observe that $\lor, \ x \lor (y \land z) \in [B]$. The transformation can be done in polynomial time by
      replacement, rewriting $\psi$ as $\bigvee_{1 \leq i \leq n} q \lor (x_i \land x'_i) \lor \bigvee_{1 \leq i \leq m} q \lor (y_i \land y'_i)$
      and using the associativity of $\lor$.
    \end{proof}
    
    \begin{theorem}\label{thm:main_variant_F}
      Let $B$ be a finite set of Boolean functions. Then, the abduction problem for propositional $B$-formulae,$\ABD(B,\MF)$,  is
      \begin{enumerate}
        \item $\SigPtwo$-complete if $\CloneS_{00}\subseteq [B]$ or $\CloneS_{10}\subseteq [B]$ or  $\CloneD_{2}\subseteq [B]$,
        \item in $\P$ and $\ParityL$-hard if $\CloneL_{2}\subseteq [B]\subseteq \CloneL$, and
        \item in $\L$ in all other cases.
      \end{enumerate}
    \end{theorem}

\section{Counting complexity} \label{sect:counting}
	The counting problem $\#\ABD(B,\MQ)$ we are interested in is the following: given an instance $\calP=(\Gamma,A,\varphi)$ of $\ABD(B,\MQ)$, compute
	$\nSols{\calP}$, the number of full explanations of $\calP$.
	
  \begin{theorem}\label{thm:counting}
    Let $B$ be a finite set of Boolean functions. Then, the abduction problem for propositional $B$-formulae, $\#\ABD(B,\MQ)$, is
    \begin{enumerate}
      \item 
      \newlength{\mylen}\setlength{\mylen}{-0.5pt}
      $\SHcoNP$-complete if $\CloneS_{02}\kern\mylen\subseteq\kern\mylen[B]$ or $\CloneS_{12}\kern\mylen\subseteq\kern\mylen [B]$ or 
      $\CloneD_{1}\kern\mylen\subseteq\kern\mylen[B]$,
      \item $\SHP$-complete if $\CloneV_{2}\subseteq [B]\subseteq \CloneM$ or $\CloneS_{10}\subseteq [B]\subseteq \CloneM$ or
      			$\CloneD_{2}\subseteq [B] \subseteq \CloneM$,
      \item in $\FP$ in all other cases.
    \end{enumerate}
  \end{theorem}
  \begin{proof} 
  
		The $\SHcoNP$-membership for $\#\ABD(B,\MQ)$ follows from the fact that checking whether a set of literals is indeed an explanation for an abduction
		problem is in $\P^\NP=\DeltaPtwo$ and from the equality $\SHclass{\DeltaPtwo}=\SHcoNP$, see \cite{hevo95}.
      
    We show the $\SHcoNP$-hardness by giving a parsimonious reduction from the following $\SHcoNP$-complete problem:
		Count the number of satisfying assignments of $\psi(x_1,\ldots,x_n)=\forall y_1\cdots \forall y_m \varphi(x_1,\ldots,x_n,y_1,\ldots,y_m)$, where
		$\varphi$ is a DNF-formula (see, \emph{e.g.}, \cite{dhk05}).
    Let $x'_1,\ldots ,x'_n,$ $ r_1,\ldots ,r_n,t,$ and $ q$ be fresh, pairwise distinct propositions. 
    We define the propositional abduction problem $\calP=(\Gamma, A, q)$ as follows:
    \begin{align*}
      \Gamma :=&\; \{x_i\rightarrow r_i, x'_i\rightarrow r_i, \neg x_i \vee \neg x_i' \mid 1\le i\le n\} \\
       \cup&\;\textstyle \{\varphi\rightarrow t\} \cup \{\bigwedge_{i=1}^n r_i\land t \rightarrow q\}, \\
      A :=&\; \{x_1,\ldots ,x_n\}\cup\{x'_1,\ldots ,x'_n\}.
    \end{align*}
    Observe that the manifestation $q$ occurs only in the formula $\bigwedge_{i=1}^n r_i\land t \rightarrow q$.
    This together with the formulae $x_i\rightarrow r_i, x'_i\rightarrow r_i, \neg x_i \vee \neg x_i'$, $1 \leq i \leq n$, enforces that every full
    explanation of $\calP$ has to select for each $i$ either $x_i$ and $\neg x'_i$, or  $\neg x_i$ and $x'_i$. By this the value of $x'_i$ is fully
    determined by the value of $x_i$ and is its dual. Moreover, it is easy to see that there is a one-to-one correspondence between the models of 
    $\psi$ and the full explanations of $\calP$.  

		Observe that since the reductions in Lemma~\ref{lem:constants_sometimes_available} are parsimonious, we can suppose w.l.o.g.\ that $B$ contains the
		two constants $\true$ and $\false$. Therefore, analogously to Proposition~\ref{prop:abd(q)-bf-sigP2} it suffices to consider the case $[B] = \CloneBF$.
		Since $\Gamma$ can be written in a normal form, it can be transformed in logarithmic space into an equivalent set of $B$-formulas. This provides a
		parsimonious reduction from the  $\SHcoNP$-complete problem we considered to $\#\ABD(B,\MQ)$.

    Note that the proposed reduction is very similar to the one proposed in \cite{hepi07}. It also proves hardness of counting the positive explanations  
    (since there is an one-to-one correspondence between full explanations and purely positive explanations), as well as 
	  hardness of counting the subset-minimal explanations, 
    (since all solutions are incomparable and hence subset-minimal).

    Let us now consider the $\SHP$-complete cases.
    When $[B]\subseteq \CloneM$, checking whether a set of literals $E$ is an explanation for an abduction problem with $B$-formulae is 
    in $\P$ (see Proposition~\ref{prop:abd(q)-m-np}). This proves membership in $\SHP$. For the hardness result, it suffices to consider the case
    $[B]=\CloneV_{2}$, because the reduction provided in
    Lemma~\ref{lem:constants_sometimes_available} is parsimonious and $\CloneV_2 \subseteq [\CloneS_{10}\cup\{1\}]$.
    We provide a Turing reduction from the problem $\#\POSTWOSAT$, which is known to be $\SHP$-complete \cite{val79}. 
    Let $\varphi =\bigwedge_{i=1}^k ( p_i\lor  q_i)$ be an instance of this problem, 
    where $p_i$ and $q_i$ are propositional variables from the set $X=\{x_1,\ldots ,x_n\}$. 
    Let $q$ be a fresh proposition. Define the propositional abduction problem $\calP=(\Gamma, A, q)$ as follows:
    \[
      \Gamma := \{p_i\lor q_i\lor q\mid 1\le i\le k  \}, \qquad A:= \{x_1,\ldots ,x_n\}.
    \]
    It is easy to check that the number of satisfying assignments for $\varphi$ is equal to $2^n-\nSols{\calP}$.
    Finally, since $\lor\in[B]=\CloneV_{2}$, 
    $\calP$ can easily be transformed in logarithmic space into an $\ABD(B,\MQ)$-instance.

    As for the tractable cases, the clones  $\CloneE$ and $\CloneN$ are easy; 
    and finally, for $[B]\subseteq \CloneL$, the number of full explanations is polynomial time computable according to \cite[Theorem~8]{hepi07}.
  \end{proof}

\section{Concluding Remarks} \label{sect:conclusion}

In this paper we provided a complete classification of the propositional abduction problem, $\ABD(B,\manif)$ for every set $B$ of allowed connectives.  
We gave results for several restrictions over the representation of manifestations. 
For instance our results show that when the knowledge base formulae are positive clauses (clone $\CloneV$) then the abduction problem is very easy (solvable in $\L$), even when the manifestations are also represented by positive clauses. But its complexity jumps to $\NP$-completeness when manifestations are represented by positive terms.
When looking at the so-called monotonic fragment (clone $\CloneM$) the abduction problem is $\NP$-complete when manifestations are represented by clauses or terms. Allowing also the manifestation to be a monotonic formulae as the knowledge base, it becomes $\SigPtwo$-complete. 
This can be intuitively explained as follows.
The complexity of the abduction rests on two sources: finding a candidate explanation and checking that it is indeed a solution. 
The $\NP$-complete cases that occur in our classification hold for problems for which verification can be performed in polynomial time. 
If both the knowledge base and the manifestation are represented by monotonic formulae, 
verifying a candidate explanation is $\co\NP$-complete.
Further our results also show that, except for the clones $\CloneL$ and $\CloneL_c$ with $c \in \{0,1,2,3\}$, all tractable cases are even trivial.

In order to get a more detailed picture of the complexity of abduction in this framework 
it would now be interesting to consider restrictions on hypotheses.
A classification in which both types of restrictions are considered would help to understand how restrictions on the representation of hypotheses and manifestations influence the complexity of the propositional abduction problem.
When restricting the hypotheses, as for instance when requiring the 
explanation to consist of positive literals only, it may happen that 
only one maximal candidate has to be considered, thus leading to 
further trivial but also some $\co\NP$-complete cases.
Also note that the upper bounds in the case of the clone $\CloneL$ 
relies on Gaussian elimination \cite{zanuttini03}.
This method fails when restricting the hypothesis to be positive. 
Determining the complexity of abduction for the clone $\CloneL$ 
when manifestations are represented by $\CloneL$-formulae 
and explanations have to be positive might hence prove to be a challenging task
(notice that this case remained unclassified in circumscriptive inference \cite{thomas09}).

\bibliographystyle{aaai}
\bibliography{thi-hannover}
\enlargethispage{0.1em}

\end{document}